\documentclass[onecolumn,a4paper,preprintnumbers,amsmath,amssymb,nofootinbib,floatfix]{revtex4}

\usepackage{color}
\usepackage{graphicx}% Include figure files
\usepackage{dcolumn}% Align table columns on decimal point
\usepackage{bm}% bold math
\usepackage{latexsym}
\usepackage{epsfig}
\usepackage{rotating}
\usepackage{bm}% bold math

\newcommand{\be}{\begin{equation}}
\newcommand{\ee}{\end{equation}}
\newcommand{\beqq}{\setlength\arraycolsep{2pt}\begin{eqnarray}}
\newcommand{\eeqq}{\vspace{0cm} \end{eqnarray}}
\newcommand{\bea}{\begin{eqnarray}}
\newcommand{\eea}{\end{eqnarray}}

\newcommand{\no}{\mbox{$n_{e_o}$}}

\newcommand{\dTo}{\mbox{$\Delta T_0$}}

\newcommand{\sigT}{\mbox{$\sigma_{\mbox{\tiny T}}$}}
\newcommand{\Tcmb}{\mbox{$T_{\mbox{\tiny CMB}}$}}

\newcommand{\Da}{\mbox{$D_{\!\mbox{\tiny A}}$}}

\begin{document}

\title{Improved constraints on violations of the Einstein equivalence principle in the electromagnetic sector with complementary cosmic probes}

\author{R. F. L. Holanda$^{1,2,3}$} \email{holandarfl@gmail.com}
\author{S. H. Pereira$^{4}$} \email{shpereira@feg.unesp.br}
\author{V. C. Busti$^{5,6}$} \email{busti@sas.upenn.edu}
\author{C. H. G. Bessa$^{7}$} \email{carlos@cosmos.phy.tufts.edu}

\affiliation{ \\$^1$Departamento de F\'{\i}sica, Universidade Federal de Sergipe, 49100-000, Aracaju - SE, Brazil,
\\ $^2$Departamento de F\'{\i}sica, Universidade Federal de Campina Grande, 58429-900, Campina Grande - PB, Brazil,\\$^3$Departamento de F\'{\i}sica, Universidade Federal do Rio Grande do Norte, 59300-000, Natal - RN, Brazil.\\$^4$Universidade Estadual Paulista (Unesp)\\Faculdade de Engenharia, Guaratinguet\'a \\ Departamento de F\'isica e Qu\'imica\\ Av. Dr. Ariberto Pereira da Cunha 333\\
12516-410 -- Guaratinguet\'a, SP, Brazil
\\$^5$Department of Physics and Astronomy, University of Pennsylvania, Philadelphia, PA 19104, USA, \\
$^6$Departamento de F\'{\i}sica Matem\'atica, Universidade de S\~ao Paulo, Rua do Mat\~ao 1371, S\~ao Paulo - SP, 05508-090, Brazil \\
$^7$Departamento de F\'{\i}sica, Universidade Federal da Para\'{\i}ba, 58051-970, Jo\~ao Pessoa - PB, Brazil,}

%\maketitle

%\keywords{Cosmology, Models beyond the standard model}

%\bigskip
\begin{abstract}
Recent results have shown that a field non-minimally coupled to the electromagnetic Lagrangian can induce a violation of the Einstein equivalence principle. { This kind of coupling is present in a very wide class of gravitation theories.} In a cosmological context, this would break the validity of the cosmic distance duality relation as well as cause a time variation of the fine structure constant. Here, we improve constraints on this scenario by using four different observables: the luminosity distance of type Ia supernovae, the angular diameter distance of galaxy clusters, the gas mass fraction of galaxy clusters and the temperature of the cosmic microwave background at different redshifts. We consider four standard parametrizations adopted in the literature and show that, due to a high complementarity of the data, the errors are shrunk between 20\% and 40\% depending on the parametrization. We also show that our constraints are weakly affected by the geometry considered to describe the galaxy clusters. In short, no violation of the Einstein equivalence principle is detected up to redshifts $\sim$ 3.   
\end{abstract}

\maketitle

%%%%%%%%%%%%%%%%%%%%%%%%%%%%%%%%%%%%%%%%%%%%%%%%%%%%%%%%%%%%%%%%%%%%%%%%%%

\section{Introduction}

Modified gravity theories have appeared recently as an alternative to General Relativity (GR) when the last one faces some difficulties to explain some observations, as the {accelerated expansion of the Universe}, galactic velocities in galaxy clusters or rotational curves of spiral galaxies. Among such new theories we can cite massive gravity theories \cite{hinterRMP}, modified Newtonian dynamic (MOND) \cite{mond}, $f(R,T)$ theories \cite{fR}, models with extra dimensions, as brane world models, Kaluza-Klein theories, {string and loop quantum cosmology theories \cite{randall,pomarol,bojowald}}. Nevertheless, some of these new theories naturally break the Einstein equivalence principle (EEP), leading to observational consequences that deserve to be tested and verified. 
%Modifications of some fundamental constants of nature, like the fine structure constant and the gravitational constant, are some effects of these generalizations. In a cosmological context, a modification {in} the expression of the luminosity distance $D_L$, which is the basis for several cosmological measurements, might also occur and must be analyzed. 

Hees et al. \cite{hees,hees1,hees2} have shown that a class of theories that explicitly breaks the EEP can be tested using recent observational data. Particularly, those theories motivated by scalar-tensor theories of gravity, {which introduce an additional coupling between the Lagrangian of the usual non-gravitational matter field with a new scalar field \cite{string,string1,klein,axion,axion1,fine, fine1,fine2,fine3,chameleon,chameleon1,chameleon2,chameleon3,fRL,BD}. In this class of theories, all the electromagnetic sector  is affected,  leading to a variation in the value of the fine structure constant ($\alpha=\alpha_0 \zeta(z)$,  where $\alpha_0$ is the current value) of the quantum electrodynamics \cite{const_alpha,const_alpha1},  a non-conservation of the photon number and, consequently,  a modification of the expression of the luminosity distance, $D_L(z)$, important for various cosmological estimates. In this context, the so-called cosmic distance duality relation, $D_L(1+z)^{-2}D_A^{-1}=1=\eta$, where $D_A$ is the angular diameter distance, as well as the  Cosmic Microwave Background (CMB) radiation temperature evolution law, $T_{CMB}(z)=T_0(1+z)$, are also affected.  {These variations are intimately and unequivocally linked (see next section).} 

Based on the results of \cite{hees,hees1,hees2}, some recent papers \cite{holandaprd,holandasaulo,holandasaulo2} have also searched, using observational data, for signatures of that class of modified gravity theories which explicitly breaks the EEP. The authors have used angular diameter distances (ADD) of galaxy clusters obtained via their X-ray surface brightness jointly { with observations of the Sunyaev-Zel'dovich effect (SZE) \cite{fil,bonamente}}, SNe Ia samples \cite{suzuki}, CMB temperature in different redshifts, $T_{CMB}(z)$ \cite{luzzi,hurier} and the most recent  X-ray gas mass fraction (GMF) data  with galaxy clusters  in the redshift range $0.078 \leq z \leq 1.063$ \cite{mantz}. {In the Ref. \cite{holandaprd} it was considered ADD + SNe Ia sample, in \cite{holandasaulo} it was used ADD + SNe Ia + $T_{CMB}$ and in the Ref. \cite{holandasaulo}, GMF + SNe Ia + $T_{CMB}$}. The crucial point in these papers is that the dependence of the SZE/X-ray technique and GMF measurements on possible departure from $\eta=\zeta=1$ was taken into account (see Section III for details). The main result found  was that no significant deviation  {for the EEP was verified by means of the electromagnetic sector}, although the results do not completely rule out those models. Thus, additional tests are still required.

In this paper, we continue searching for deviations of the EEP by considering several cosmological observations and the class of models that explicitly breaks the EEP in the electromagnetic sector discussed in \cite{hees,hees1,hees2} . We consider a more complete analysis, including ADD + GMF + SNe Ia + $T_{CMB}$ data. However, in our analyses the GMF measurements are used by using two methods: in the method I we use GMF measurements obtained separately via X-ray and SZE observations for a same galaxy cluster and in the method II the X-ray GMF observations of galaxy clusters are used jointly with SNe Ia data. Therefore, it is important to emphasize that we not only combine previous tests but add one more: the method I.  As result, this more comprehensive analysis due to a larger data set allowed us to decrease the errors roughly by 20\% to 40\% depending on the adopted functional form for the deviation. Once more we show that no significant deviation  {of the EEP} is verified.

This paper is organized as follows: Section II we briefly revise the cosmological equations for a class of scalar-tensor theories based on \cite{hees1}. The consequences for cosmological measurements are presented in Section III. The cosmological data are in Section IV, and the analises and results in Section V. We finish with a conclusion in Section VI. 

\section{Scalar-tensor theories coupled to electromagnetic sector}

{A specific class of modified gravity theories characterized by a universal non-minimal coupling between an extra scalar field $\Phi$ to gravity was studied recently by Hees et al. \cite{hees,hees1,hees2}. In such scalar-tensor
theories the standard matter Lagrangian $\mathcal{L}_i$ and the scalar field $\Phi$ are represented by the action:}
\begin{eqnarray}
S= \int  d^4x \sqrt{-g} \Big[ f_i(\Phi) \mathcal{L}_i (g_{\mu \nu}, \Psi_i) +  \frac{1}{2\kappa}\left(\Phi  R-\frac{\omega(\Phi)}{\Phi} (\partial_\sigma \Phi)^2-V(\Phi) \right)\Big], \label{action}
\end{eqnarray}
where $R$ is the Ricci scalar for the metric $g_{\mu \nu}$ with determinant $g$, $\kappa=8\pi G$, where $G$ is the gravitational constant, $V(\Phi)$ is the scalar-field potential, $f_i(\Phi)$ and $\omega(\Phi)$ are arbitrary functions of $\Phi$.  $\mathcal{L}_i$ is the matter Lagrangian for the non-gravitational fields $\Psi_i$, where for a matter content consisting of a perfect fluid, for instance, we have $\mathcal{L}_{Mat}(g_{\mu \nu}, \Psi_{Mat})$, where $\Psi_{Mat}$ stands for the field describing the perfect fluid. For the electromagnetic radiation we have $\mathcal{L}_{EM}(g_{\mu \nu}, \Psi_{EM})$, where $\Psi_{EM}=A^\mu$ stands for the 4-vector potential. {From the extremization of the action (\ref{action}) follows the Einstein field equations \cite{hees1}:
\begin{equation}
R_{\mu\nu}-{1\over 2}g_{\mu\nu}R=\kappa \frac{f_i(\Phi)}{\Phi}T^i_{\mu\nu}+\frac{1}{\Phi}[\nabla_\mu \nabla_\nu - g_{\mu\nu}\square]\Phi+ \frac{\omega(\Phi)}{\Phi^2}\bigg[\partial_\mu\Phi \partial_\nu\Phi-{1\over 2}g_{\mu\nu}(\partial_\alpha\Phi)^2\bigg]-g_{\mu\nu}\frac{V(\Phi)}{2\Phi}\,,
\end{equation}
where the stress-energy tensor is given by $T^i_{\mu\nu}=(-2/\sqrt{-g})\delta(\sqrt{-g}\mathcal{L}_i)/\delta g^{\mu\nu}$. It is clear that the cases $f_i(\Phi)\neq 1$ and/or $\Phi\neq 1$ will represent  the break of EEP, while} the limit $\Phi\to 1$, $f_i(\Phi)\to 1$, $\omega(\Phi)=0$ and $V(\Phi)=0$ corresponds to the standard framework, for some matter Lagrangian. The case $\omega(\Phi)= constant$ and $f_i(\Phi)=1$ stands for the Brans-Dicke theory \cite{BD}. The dilaton \cite{string} and pressuron theory \cite{hees3} also follows from that action.

{In order to study just the break of EEP due to a coupling of a single scalar field $\Phi$ with the electromagnetic sector of theory, which is motivated by the first term of the action (1) or the first term on the right side of (2), we just need to consider the usual electromagnetic Lagrangian coupled to $\Phi$ through $f_i(\Phi)$.} In which follows we will present the main results in the case where the electromagnetic field is the only matter field present into the action (\ref{action}), although it is not a significant source of curvature (photons are just test particles non-minimally coupled to $\Phi$). In the vacuum  the Lagrangian is \cite{hees2}
\begin{equation}
\mathcal{L}_{EM}(g_{\mu \nu}, A^\mu)=-\frac{1}{4}F^{\mu\nu}F_{\mu\nu}\,,
\end{equation}
with $F^{\mu\nu}=\partial^\mu A^\nu - \partial^\nu A^\mu$  {and we will consider a coupling $f_i(\Phi)=f_{EM}(\Phi)$.} Variation of the action with respect to the 4-potential $A^\mu$ gives the modified Maxwell equations
\begin{equation}
\nabla_\nu \left(f_{EM}(\Phi)F^{\mu\nu}\right)=0.
\end{equation}
Following the standard procedure in GR \cite{minazzoli2,EMM}, we expand the 4-potential as $A^\mu = \Re \left\{ \left(b^\mu + \epsilon c^\mu + O(\epsilon^2) \right) \exp^{i \theta / \epsilon} \right\}$ and use the Lorenz gauge which { leads} to the usual null-geodesic at leading order. The next order of the modified Maxwell equations is given by
\begin{eqnarray}
        k^\nu \nabla_\nu b &=&-\frac{1}{2}b\nabla_\nu k^\nu -\frac{1}{2}bk^\nu \partial_\nu \ln f_{EM}(\Phi) \label{amplitude}\\
        k^\nu \nabla_\nu h^\mu &=&\frac{1}{2}k^\mu h^\nu\partial_\nu \ln f_{EM}(\Phi)
\end{eqnarray}    
where $b$ is the amplitude of $b^\mu=b h^\mu$, $h^\mu$ is the polarisation vector and $k_\mu \equiv \partial_\mu \theta$. The conservation law of the number of photons is written as
\begin{equation}
    \nabla_\nu \left(b^2k^\nu\right)=-b^2 k^\nu\partial_\nu \ln f_{EM}(\Phi).
\end{equation}
 {The wave vector in the flat FRW metric in spherical coordinate is $k^\mu=(k^0,k^r,0,0)=(-k_0,k_0/a(t),0,0)$} and it can be showed that the quantity $K=b(t,r) r a(t) \sqrt{f_{EM}(\Phi(t))}$ is constant along a geodesic. 

The flux of energy comes from the $T^{0i}$ component of the energy momentum tensor and is given by
\begin{equation}\label{flux_tmp}
 F_0 =\left|a_0 b^2 k^0 k^r\right| =\frac{k_r^2b^2 }{a^2_0}=\frac{k_r^2 K^2}{r_0^2 a^4_0f_{EM}(\Phi_0)}=\frac{C}{r_0^2a_0^4f_{EM}(\Phi_0)}
\end{equation}
where $C$ is a constant. The emitted flux is
\begin{equation}
F_e=\frac{C}{r_e^2a_e^4f_{EM}(\Phi_e)}
\end{equation}
where the index $e$ refers to the emitted signal. The angular integral of this defines the luminosity $L_e$
\begin{equation}
L_e=\frac{4\pi C}{a_e^2f_{EM}(\Phi_e)}.
\end{equation}
Finally, the expression for the distance of luminosity is
\begin{equation}
D_L=\left(\frac{L_e}{4\pi F_0}\right)^{1/2}=\frac{a_0}{a_e}a_0 r_0 \sqrt{\frac{f_{EM}(\Phi_0)}{f_{EM}(\Phi_e)}}=c(1+z)\sqrt{\frac{f_{EM}(\Phi_0)}{f_{EM}(\Phi(z))}}\int_0^z\frac{dz}{H(z)}.\label{DL}
\end{equation}
Such expression clearly shows that $D_L$ is slightly modified for a non-minimal coupling $f_{EM}$ between the electromagnetic Lagrangian and an extra scalar field.

On the other hand, the angular diameter distance $D_A$ is {a purely geometric quantity that is the same as in ordinary electromagnetism}
\begin{equation}
D_A(z)=\frac{c}{(1+z)} \int_{0}^{z} \frac{dz'}{H(z')}\,.\label{DA}
\end{equation}
By comparing with (\ref{DL}) we have:
\begin{equation}
\frac{D_L(z)}{D_A(z)(1+z)^2}= \sqrt{\frac{f_{EM}(\Phi_0)}{f_{EM}(\Phi(z))}}\equiv\eta (z)\,,\label{DLDA}
\end{equation}
where we have defined the parameter $\eta(z)$ related to $f_{EM}(\Phi(z))$ for convenience, when $\eta(z)=1$, the above relation is also known as the cosmic distance duality relation (CDDR).  { The CDDR is a relation between angular diameter and luminosity distances for a given redshift, $z$, namely, $ D_L D_A^{-1} (1+z)^{-2}=1$. This equation is an astronomical consequence deduced from the reciprocity theorem when photons follow null geodesics, the geodesic deviation equation is valid and  the number of photons is conserved. It plays an essential role in cosmological observations and  in the last years it has been tested by several authors in different cosmological context \cite{distance,distance1,distance2,distance3,distance4,distance5} (see Table I in \cite{hbajcap} for recent results).} 

 {As commented earlier, the kind of coupling explored in this paper leads to a variation in the value of the fine structure constant, $\alpha=\alpha_0 \zeta(z)$, violations of the cosmic distance duality relation, $D_L(1+z)^{-2}D_A^{-1}=\eta(z)$,  as well as a modification in the CMB temperature evolution law, $T_{CMB}(z)=T_0(1+z)^{1-\tau}$, and these possible variations are intimately and unequivocally linked. As shown in Ref. \cite{hees} (see their equations (12) and (34)), if one parametrizes a possible departure from the CDDR validity with a $\eta(z)$ term, the consequent deviation in the  CMB temperature evolution law and the temporal evolution of the fine structure constant have to be described by
\begin{equation}
\label{alpha}
\frac{\Delta \alpha}{\alpha}= \zeta(z) -1 = \eta^2 (z) -1
\end{equation}
and
\begin{equation}
T(z)=T_0(1+z)[0.88+0.12 \eta^2(z)]. \label{T}
\end{equation} }

%For completeness only, as shown by \cite{string}, the variation of the fine-structure constant $\alpha$ with time is related to the coupling $f_{Mat}$ of the scalar field $\Phi$ to the matter part of the Lagrangean, such that $\alpha \propto f_{Mat}^{-1}$ and 
%\begin{equation}
%\frac{\dot{\alpha}}{\alpha}= - \left.\frac{\dot{f}_{Mat}}{f_{Mat}}\right|_0= -n \left.\frac{\dot{\Phi}}{\Phi}\right|_0,
%\end{equation}
%where $n=1/2$ for the pressuron theory.

Deviation from such relations will have several consequences on cosmological observables, as modifications of angular diameter distances of galaxy clusters obtained via their X-ray and SZE observations, the gas mass fraction of galaxy clusters and the CMB temperature evolution law. Thus we have a robust method to test the break of the equivalence principle in the electromagnetic sector.

In this work, in order to better explore possible break of EEP, we consider four widely used parametrizations for the $\eta(z)$ function:

\begin{itemize} 

\item P1: $\eta(z)=1+\eta_0 z$

\item P2: $\eta(z)=1+\eta_0 z/(1+z)$

\item P3: $\eta(z)=(1+z)^{\eta_0}$

\item P4: $\eta(z)=1+ \eta_0 \ln(1+z)$ 
\end{itemize}
where  $\eta_0$ is the parameter to be constrained and the limit  $\eta_0=0$ (or $\eta(z)=1$)  {corresponds to no violation of the EEP}.

\section{Consequences for cosmological measurements}

\begin{figure*}[t]
\centering
\includegraphics[width=0.47\textwidth]{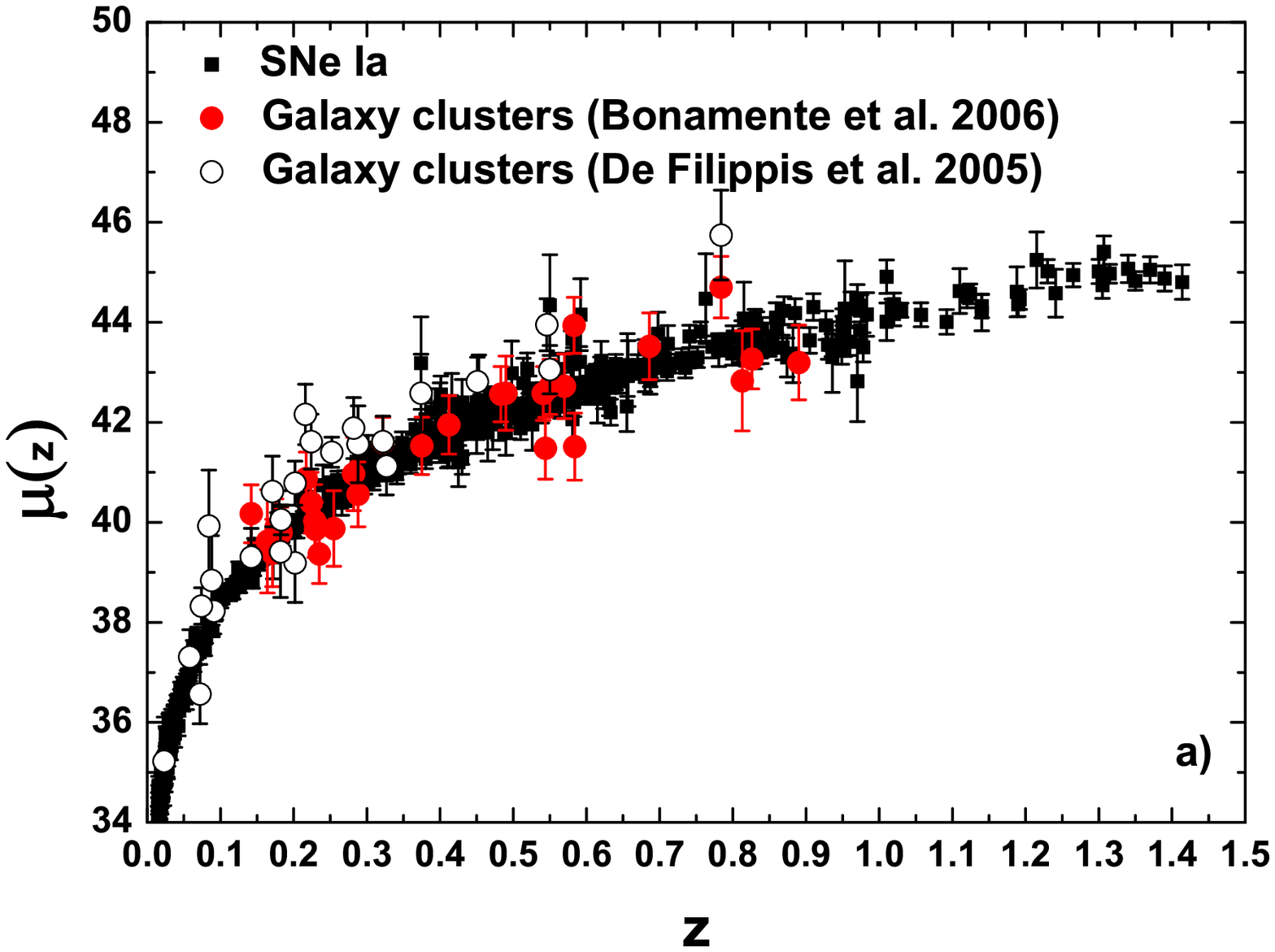}
\hspace{0.3cm}
\includegraphics[width=0.47\textwidth]{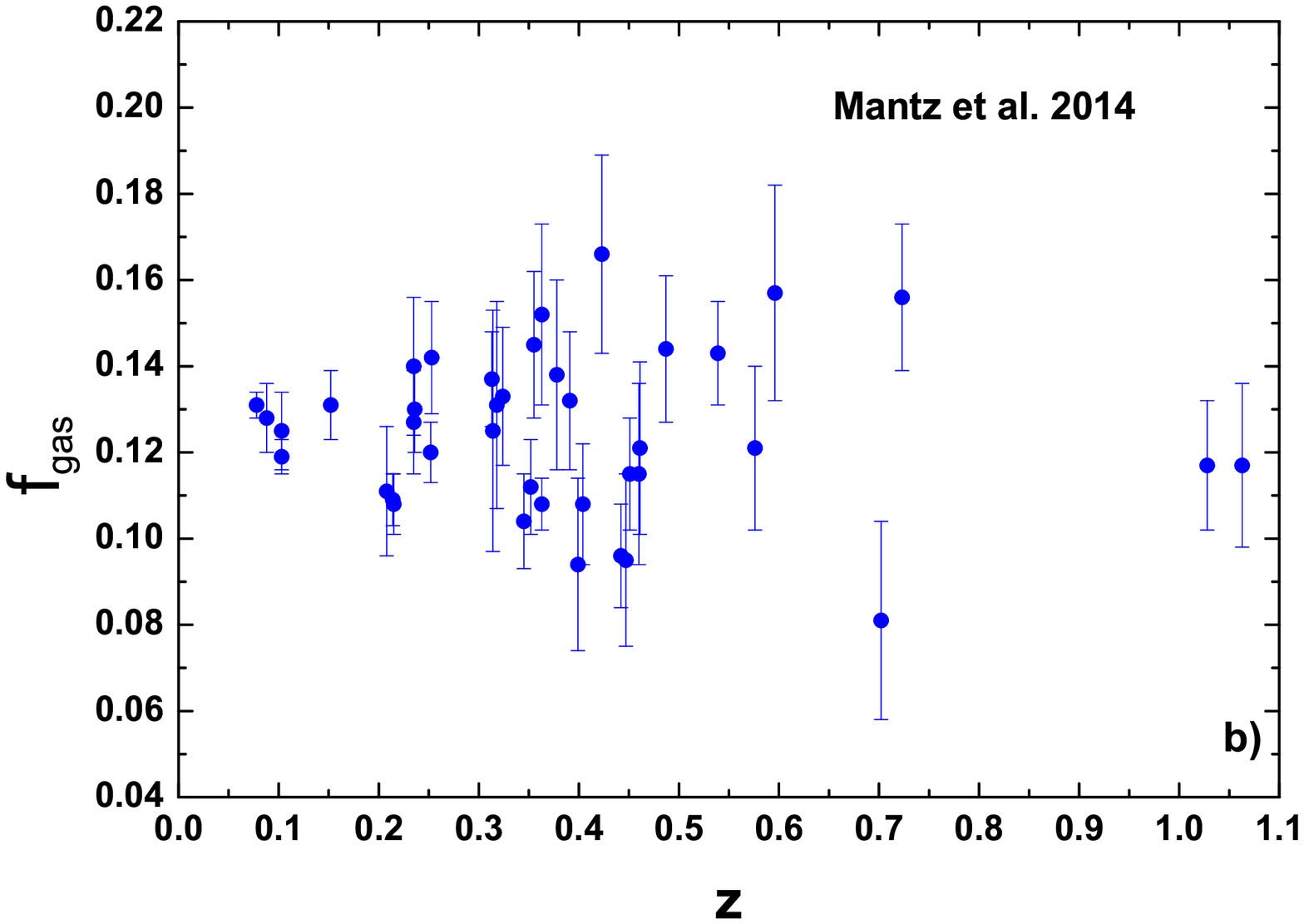}
\caption{ In Fig.(a) we plot the 580 distance moduli (black squares) of SNe Ia \cite{suzuki}, 29 (red circle) and 25 (open circle) distance moduli of galaxy clusters from Refs.\cite{bonamente,fil}, respectively. In Fig.(b) we plot the 40 gas mass fraction (GMF) data \cite{mantz}.  }
\end{figure*}

In the following subsections, we discuss the consequences of the EEP breaking on cosmological measurements and explain the basic equations used in our analyses.

\subsection{Angular diameter distance of galaxy clusters}

 {The angular diameter distance of galaxy clusters can be obtained from their Sunyaev-Zel'dovich effect (SZE) \cite{sunyaev} and X-ray observations \cite{bonamente,de}. In this point, it is important to discuss the key points of this technique. The SZE is a small distortion of the CMB spectrum, due to the inverse Compton scattering of the CMB photons passing through a population of hot electrons in intra-cluster medium. The  temperature decrement in Rayleigh-Jeans portion of CMB radiation spectrum that crosses the center of the cluster is given by\footnote{For simplicity, we assume the spherical $\beta$-model to the galaxy clusters \cite{cavaliere}, where the electron density of the hot intra-cluster gas has a profile of the form: $n_e(r) =n_{e0}\left[1+\left(\frac{r}{r_c}\right)^2\right]^{-3\beta/2}$.}
\begin{equation}
\label{eqsze3} \Delta T_0 \equiv T_0
f(\nu, T_{\rm e}) \frac{ \sigma_{\rm T} k_{\rm B} T_{\rm e}}{m_{\rm e}
c^2}n_{e0} \sqrt{\pi} \theta_c D_A g\left(\beta/2\right),
\end{equation}
with
\begin{equation}
g(\alpha)\equiv\frac{\Gamma \left[3\alpha-1/2\right]}{\Gamma \left[3
\alpha\right]}, \label{galfa}
\end{equation}
where $\Gamma(\alpha)$ is the gamma function, $\theta_c$ is the cluster core angular size, $n_{e0}$ is the central electronic density of the intra-cluster medium, $k_B$ the Boltzmann constant, $T_e$ is the electronic temperature, $T_0$ = 2.728K is the present-day temperature of the CMB,  $f(\nu, T_e)$ accounts for frequency shift and relativistic corrections \cite{itoh}, and $m_e$ the electron mass. The Thompson cross section, $\sigma_{\rm T}$, can be written in terms of the fine structure constant ($\alpha=e^2/c \hbar$) as \cite{galli}
\begin{equation}
\label{thom}
\sigma_{\rm T}=\frac{8\pi \hbar^2 }{3 m_e^2 c^2}\alpha^2,
\end{equation}
where $e$ is the electronic charge,  $c$ is the speed of light and $\hbar$ is the reduced Planck constant.}

 {On the other hand, the X-ray emission is due to thermal bremsstrahlung and the central surface brightness is given by \cite{uza}
\begin{equation}
\label{eqsxb2} S_{X0} \equiv \frac{D_A^2 \Lambda_{e} }{D_L^2 4 \sqrt{\pi} } n_{e0}^2 \theta_c D_A \ g(\beta),
\end{equation}
which clearly depends on the CDDR \cite{uza}. The term $\Lambda_{e}$ is the central X-ray cooling function of the intra-cluster
medium \cite{sarasin}. Thus, the angular diameter distance of a galaxy cluster can be found by eliminating $n_{e0}$ in the Eqs. (\ref{eqsze3}) and (\ref{eqsxb2}), taking the validity of the CDDR and considering the constancy of $\alpha$. However, if one considers  $\alpha=\alpha_0\zeta(z)$ and $D_L D_A^{-1} (1+z)^{-2}=\eta(z) \neq 1$, a more general result appears \cite{uza,colaco}}
 {
\begin{widetext}
\begin{eqnarray}
\label{eqobl7}
{{D}}_A(z) &= & \left[ \frac{\Delta {T_0}^2}{S_{\rm X0}}
\left( \frac{m_{\rm e} c^2}{k_{\rm B} T_{e} } \right)^2
\frac{g\left(\beta\right)}{g(\beta/2)^2\ \theta_{\rm c}}
\right] \times %\nonumber \\
%& & \times
 \left[ \frac{\Lambda_e \zeta^3(z)}{4 \pi^{3/2}f(\nu,T_{\rm
e})^2\ {(T_0)}^2 {\sigma_{\rm T}}^2\ (1+z)^4}\frac{1}{\zeta^4(z)\eta(z)^2} \right]. \nonumber \\
\end{eqnarray}
\end{widetext}
The observational quantity in the above equation is}
 {
\begin{widetext}
\begin{eqnarray}
\label{eqobl7}
D^{\: data}_A(z) &= & \left[ \frac{\Delta {T_0}^2}{S_{\rm X0}}
\left( \frac{m_{\rm e} c^2}{k_{\rm B} T_{e} } \right)^2
\frac{g\left(\beta\right)}{g(\beta/2)^2\ \theta_{\rm c}}
\right] \times %\nonumber \\
%& & \times
 \left[ \frac{\Lambda_e }{4 \pi^{3/2}f(\nu,T_{\rm
e})^2\ {(T_0)}^2 {\sigma_{\rm T}}^2\ (1+z)^4}\right].\nonumber \\
\end{eqnarray}
\end{widetext}
So,  the currently measured quantity is $D_A^{\: data}(z)=D_A(z) \; \eta^2(z) \zeta(z)$, where $D_A(z)$ is the true angular diameter distance.} In this way, by considering the large class of theories proposed by \cite{hees,hees2} where the relation (14) is valid,  we have access to \cite{holandaprd}

\begin{equation}
\frac{D_L}{(1+z)^2D_A^{data}(z)}=\eta^{-3}(z).
\end{equation}
By using the equation above, we define the distance modulus of a galaxy cluster (GC) data as
\begin{equation}
\label{muda}
 \mu_{GC}(\eta ,z)=5\log[\eta^{-3}(z)D^{data}_A(z)(1+z)^2]+25. 
\end{equation}
Thus, if we have SNe Ia distance modulus measurements, $\mu(z)$, at identical redshifts of galaxy clusters, we can put observational constraints on the $\eta$ parameter.}

\subsection{Gas mass fraction of galaxy clusters}

Here, we can put limits on $\eta(z)$ from two methods:

\begin{figure*}[t]
\centering
\includegraphics[width=0.47\textwidth]{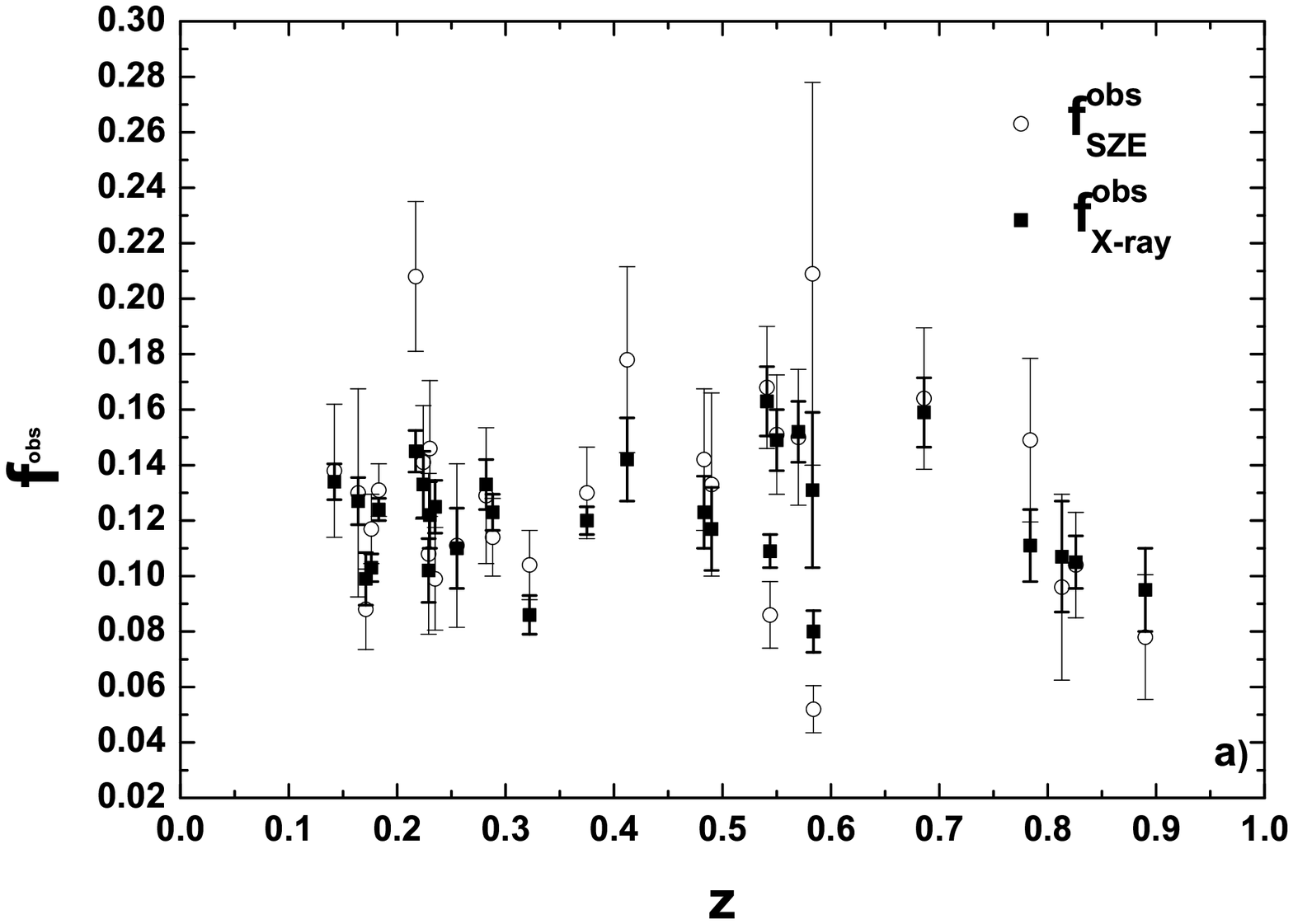}
\hspace{0.3cm}
\includegraphics[width=0.47\textwidth]{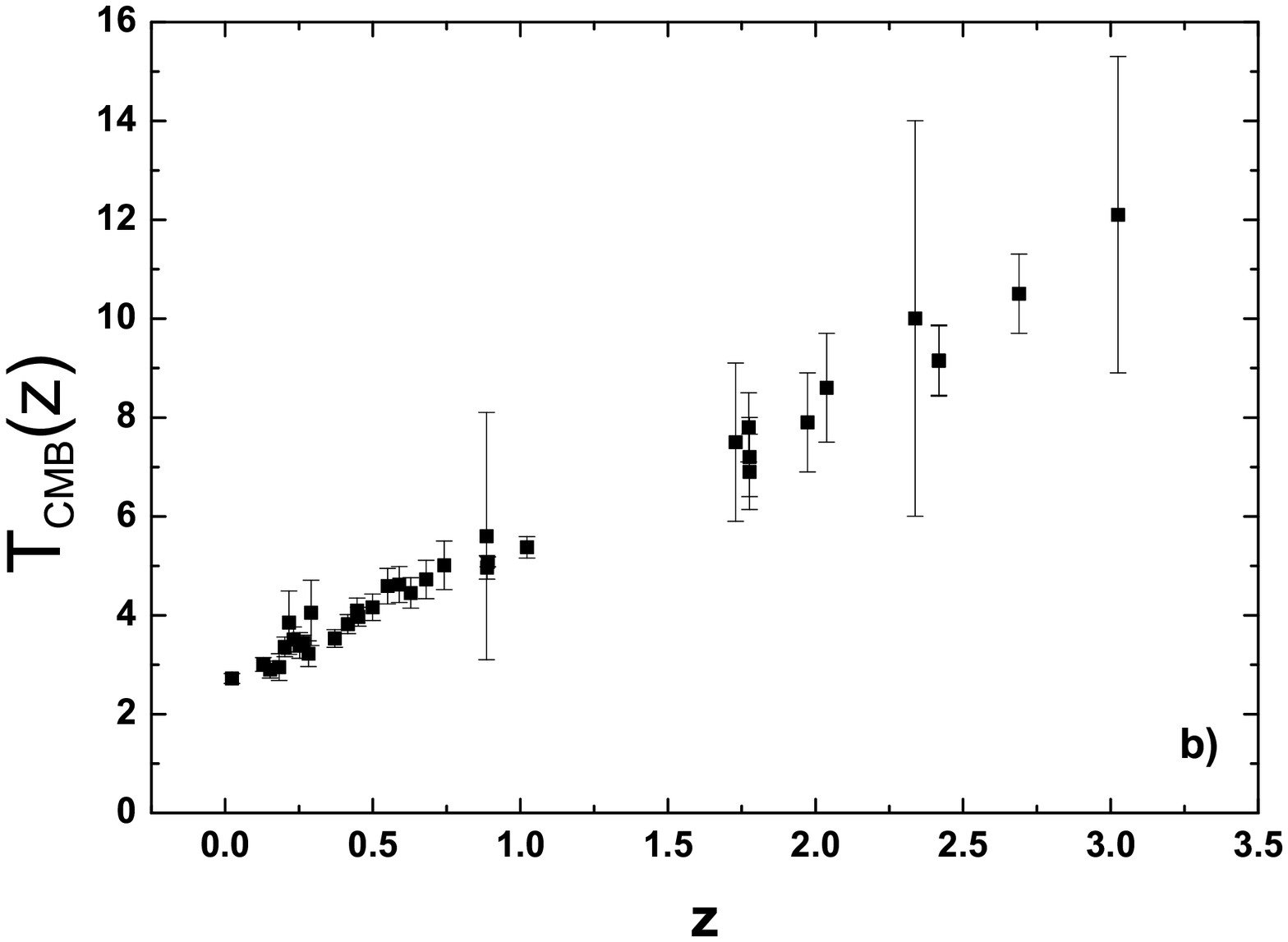}
\caption{ In Fig.(a) we plot the 29 GMF from Ref. \cite{laroque}. In Fig.(b) we plot the 36 $T_{CMB}(z)$ data \cite{luzzi,hurier}.}
\end{figure*}

\subsubsection{Method I}

 {The gas mass fraction is defined as \cite{sasaki}
\begin{equation}
 f_{gas}=\frac{M_{gas}}{M_{tot}},
 \label{eq3.14}
\end{equation}
where $M_{tot}$ is the total mass  and  $M_{gas}$ is the gas mass obtained by  integrating the gas density model, for instance,  the spherical $\beta$-model. Under the hydrostatic equilibrium, isothermality and the spherical $\beta$-model assumptions, the $M_{tot}$ within a $R$ radius is given by \cite{grego}
\begin{eqnarray}
 M_{tot}(<R)=\frac{3\beta k_{\rm B}T_e}{\mu_m Gm_H}\left[\frac{R^3}{(r_{c}^{2}+R^2)}\right],
 \label{Mtot}
\end{eqnarray}
where $\mu_m$ and $m_H$ are, respectively, the  total mean molecular weight and the proton mass, $G$ is the gravitational constant and $r_c$ is the  cluster core radius. On the other hand, the $M_{gas}$ within a volume $V$ is obtained by 
\begin{eqnarray}
 M_{gas}(<V)=\frac{8\pi n_{e0}m_Hr_{c}^{3}}{(1+X)}I_M(y,\beta),
 \label{Mgas}
\end{eqnarray}
where
\begin{eqnarray}
 I_M(R/r_c,\beta)\equiv \int_{0}^{R/r_c}{(1+x^2)^{-\frac{3\beta}{2}}x^2dx}\;,
 \label{eq3.10}
\end{eqnarray}
$x=r/r_c$ and  $X$ is the hydrogen abundance. In this way, the gas mass fraction of a galaxy cluster is
\begin{eqnarray}
f_{gas}=\frac{8\pi m_H^2\mu_m G n_{e0}}{3(1+X)\beta k_BT_G}\left[\frac{(r_{c}^{5}+r_c^3R^2)}{R^3}\right]I_M({R \over r_c},\beta).
\label{fgas} 
\end{eqnarray}
The quantity $n_ {e0}$ in the above equation can be determined from two different kinds of observations: X-rays surface brightness and the Sunyaev-Zeldovich effect.}

 {By Using SZE observations, the central electron density can  be expressed as \cite{laroque}
\begin{equation}
n_{e0}^{SZE} =  \frac{\dTo \,m_e c^2 \:\Gamma(\frac{3}{2}\beta)}{f_{(\nu, T_e)}
  \Tcmb \sigT\, k_{\rm B} T_e \Da \pi^{1/2} \:\Gamma(\frac{3}{2}\beta -
  \frac{1}{2})\, \theta_c}.
\label{eq:sz_ne0}
\end{equation}
From the Eq. (\ref{thom}), one may show that the current gas mass fraction measurements via SZE depend on $\alpha$ as \cite{suzana}  
\begin{equation}
f^{obs}_{SZE} \propto {\alpha^{-2} }\;,
\end{equation}
or still, if $\alpha=\alpha_0 \zeta(z)$, 
\begin{equation}
f_{SZE}^{th} \propto  [\zeta(z)]^{-2} \;.
\end{equation}
}
 {On the other hand, from X-ray observations, the bolometric luminosity is given by \cite{sarasin} 
\begin{eqnarray}
 L_x&=&\left(\frac{2\pi k_BT_e}{3m_e}\right)^{\frac{1}{2}}\frac{2^4e^6}{3\hbar m_ec^3}g_B(T_e)\frac{2}{(1+X)}4\pi n_{e0} \int_{0}^{R}{\left(1+\frac{r^2}{r_{c}^{2}}\right)^{-3\beta}r^2dr}.
 \label{eq3.19}
\end{eqnarray}
Defining
\begin{eqnarray}
I_L(R/r_c,\beta)\equiv \int_{0}^{R/r_c}{(1+x^2)^{-3\beta}x^2dx}\;,
\label{eq3.20}
\end{eqnarray}
we obtain the equation for the bolometric luminosity
\begin{eqnarray}
 L_x&=&\left(\frac{2\pi k_BT_e}{3m_e}\right)^{\frac{1}{2}}\frac{2^4e^6}{3\hbar m_ec^3}g_B(T_e)\frac{2}{(1+X)}4\pi n_{e0}^2r_{c}^{3} I_L(R/r_c,\beta),
 \label{eq3.21}
\end{eqnarray}
which can be rewritten in terms of $\alpha$ as 
\begin{eqnarray}
 L_x&=&\alpha^3\left(\frac{2\pi k_BT_e}{3m_e}\right)^{\frac{1}{2}}\frac{2^4\hbar^2}{3 m_e}g_B(T_e)\frac{2}{(1+X)}4\pi n_{e0}^2 D^2_A \theta^2_c r_{c} I_L(R/r_c,\beta)\;,
 \label{eq3.22}
\end{eqnarray}
{where  $D_A$ is the angular diameter distance, $n_e$ is the electronic density of gas, $g_B$ is the Gaunt factor which takes into account the corrections due quantum and relativistic  effects of Bremsstrahlung emission. However, the quantity $L_x$, the total X-ray energy per second leaving the galaxy cluster, is not an observable. The quantity observable is the X-ray flux
\begin{equation}
 F^x = L_x/4\pi D^2_L,
\label{fluxo}
\end{equation}
where $D_L$ is the luminosity distance. Thus, as one may see from equations (\ref{eq3.22}) and (\ref{fluxo}), $n_{e0}$  is  $\propto \alpha^{-3/2} D_L/D_A$. Therefore, if $\alpha=\alpha_0\zeta(z)$ and the cosmic distance duality relation is $D_L (1+z)^{-2}/D_A=\eta(z)$, the gas mass fraction measurements  extracted from X-ray data are affected by a possible departure of $\alpha_0$ and $\eta(z) =1$, such as \cite{hra,suzana}}
\begin{equation}
f_{X-ray}^{th} \propto  [\zeta(z)]^{-3/2} \eta(z) \;.
\end{equation}
As discussed in \cite{suzana}, current $f^{obs}_{X-ray}$ and $f^{obs}_{SZE}$ measurements have been obtained by assuming $\zeta(z)=1$ and $\eta(z)=1$. However, if $\alpha$ varies over the cosmic time, the real gas mass fraction from X-ray ($f_{X-ray}^{th}$) and SZE ($f_{SZE}^{th}$) observations  should be related with the current  observations by 
\begin{equation}
f_{X-ray}^{th}=\zeta(z)^{-3/2} \eta(z) f^{obs}_{X-ray}\;,
\end{equation}
\begin{equation}
f_{SZE}^{th}=\zeta(z)^{-2} f^{obs}_{SZE}\;. 
\end{equation}
In this way, as one would expect, $f_{gas}$ measurements from both techniques have to agree with each other since they are measuring the very same quantity ($f_{X-ray}^{th} = f_{SZE}^{th}$). Thus,  the expression relating current X-ray  and  SZE gas mass fraction observations is given by:
\begin{equation} 
\label{relation}
f^{obs}_{SZE}=\zeta(z)^{1/2} \eta(z) f^{obs}_{X-ray}\;.
\end{equation}
Therefore, by using Eq. (14) \cite{hees,hees2},  we have access to
\begin{equation}
\label{fz}
f^{obs}_{SZE}(z)=\eta(z)^2f^{obs}_{X-ray}.
\end{equation}
If one has $f^{obs}_{SZE}$ and $f^{obs}_{X-ray}$ for the same galaxy cluster it is possible to impose limits on $\eta(z)$.}

\subsubsection{Method II}

 {Since galaxy clusters  are the largest virialized objects in the Universe, one may expect that their cluster baryon fraction is a faithful representation of the cosmological average baryon fraction $\Omega_b/\Omega_M$, in which $\Omega_b$ and $\Omega_M$ are, respectively, the fractional mass density of baryons and all matter.} Thus, X-ray observations of galaxy clusters can  be used to constrain cosmological parameters if one assumes that $f_{gas}$  is the same at all $z$ \cite{sasaki}. For this context,  X-ray gas mass fraction observations of galaxy clusters are used to constrain cosmological parameters from an expression \cite{ale,ale2,ale3} that depends on the CDDR, such as \cite{gon}:
\begin{equation}
\label{GasFrac}
f^{obs}_{X-ray}(z)=N\left[\frac{D_L^* D_A^{*1/2}}{D_L D_A^{1/2}}\right],
\end{equation}
where the symbol * denotes the  quantities  from a fiducial cosmological model (usually a flat $\Lambda$CDM model where $\eta=1$) that are used in the observations and the normalization factor $N$ carries all the information about the matter content in the cluster. In the Ref. \cite{gon}, the authors showed that this quantity is affected by a possible departure from $\eta =1$ and the Eq. (\ref{GasFrac}) must be rewritten as
\begin{eqnarray}
\label{GasFrac3}
f^{obs}_{X-ray}(z) &=& N \left[\frac{\eta^{1/2}(z)D_L^{*3/2}}{D_{L}^{3/2}}\right],
\end{eqnarray}
where the $\eta(z)$ parameter appears after using $ D_L D_A^{-1} (1+z)^{-2}=\eta(z)$. 

However, as discussed earlier, the Ref. \cite{suzana} showed that the gas mass fraction measurements  extracted from X-ray data are also affected by a possible departure of $\zeta(z)=1$, such as 
\begin{equation}
f_{X-ray}^{th} \propto  [\zeta(z)]^{-3/2}
\end{equation}  
or, by considering the Eq. (\ref{alpha}), 
\begin{equation}
f_{X-ray}^{th} \propto  \eta(z)^{-3}.
\end{equation}
In this context, the quantity $f^{obs}_{X-ray}$  may still deviate from  its true value by a factor $\eta^{-3}$, which does not have a counterpart on the right side of the Eq. (\ref{GasFrac3}). Then, this expression has to be modified to \cite{holandasaulo2} 
\begin{eqnarray}
\label{GasFrac4}
f^{obs}_{X-ray}(z) &=& N \left[\frac{\eta^{7/2}(z)D_L^{*3/2}}{D_L^{3/2}}\right].
\end{eqnarray}
Finally, the luminosity distance of a galaxy cluster can be obtained from its gas mass fraction  by 
\begin{eqnarray}
\label{dl}
D_L &=& \eta^{7/3}(z)D_L^{*}[N/f^{obs}_{X-ray}(z)]^{2/3},
\end{eqnarray}
and so, its distance modulus is
\begin{equation}
\label{muf}
 \mu_{GC}(\eta ,z)=5\log[\eta^{7/3}(z)D_L{*}[N/f^{obs}_{X-ray}(z)]^{2/3}]+25. 
\end{equation} 
Again, if we have SNe Ia distance module measurements, $\mu(z)$, at identical redshifts of galaxy clusters, we can put observational constraints on the $\eta(z)$ parameter.

\subsection{CMB temperature evolution law}

The last and more simple modified equation is the CMB temperature evolution law $T_{CMB}(z)$. According to the
 Refs. \cite{hees,hees2}, the standard CMB temperature evolution law, $T_{CMB}(z)=T_0(1+z)$, has been modified to  

\begin{equation}
T_{CMB}(z)=T_0(1+z)\left[1+0.12\eta(z)^2\right],
\label{var}
\end{equation}
 {if  one considers violations of the cosmic distance duality relation such as $D_L(1+z)^{-2}D_A^{-1}=\eta(z)$.}

\section{Cosmological data}

In our analysis, we use the following data set:

\begin{itemize}

\item 29 X-ray and SZE gas mass fraction measurements as given in  Ref. \cite{laroque}. Actually, the sample consists of 38 massive galaxy clusters spanning redshifts from 0.14 up to 0.89. In order to perform a realistic model for the cluster gas distribution, the gas density
was modeled with the non-isothermal double $\beta$-model that generalizes the single $\beta$-model profile. An important aspect concerning the galaxy cluster sample shown in Fig. 1c is that some objects are not well described by the  hydrostatic equilibrium model (see Table
6 in Ref. \cite{bonamente}). They are: Abell 665, ZW 3146, RX J1347.5-1145, MS 1358.4 + 6245, Abell 1835, MACS J1423+2404, Abell 1914, Abell 2163, Abell 2204. By excluding these objects from our sample, we end up with a subsample of 29 galaxy clusters. Moreover, it is worth mentioning that the shape parameters of the gas density model ($\theta_c$ and $\beta$) were obtained from a joint analysis of the X-ray and SZE data, which makes the SZE gas mass fraction not independent\footnote{In all data with SZE observations used in our analysis the frequency used to obtain the SZE signal in galaxy clusters sample considered was 30 GHz, in this band the effect on the SZE from a variation of $T_{CMB}$ is completely negligible \cite{mel}. Therefore, we do not consider a modified CMB temperature evolution law in the galaxy cluster data.}. However, simulations have shown that the values of $\theta_c$ and $\beta$ computed separately by SZE and X-ray observations agree at 1$\sigma$ level within a radius $r_{2500}$, the same used in the La Roque et al. \cite{laroque} observations. 

\item Two samples of angular diameter distance of galaxy clusters obtained via their SZE+X-ray observations. These samples are different from each other by the assumptions used to describe the clusters (see Fig. 1a). The first one corresponds to 29 angular diameter distances of galaxy clusters compiled by Ref.  \cite{bonamente}. The 29 galaxy clusters here are identical to those in the previous item, where the gas density was also modeled with the non-isothermal double $\beta$-model. The second one is that presented by the Ref. \cite{fil}, where the X-ray surface brightness was described by an elliptical isothermal $\beta$-model. In this case, the galaxy clusters are distributed over the redshift interval $0.023 \leq z \leq 0.784$. It is critical to consider different assumptions on the galaxy clusters morphology since the distance depends on the hypotheses considered. In both samples, we have added a conservative 12\% of systematic error (see Table 3 in \cite{bonamente}). 

\item The most recent X-ray mass fraction measurements of 40 galaxy clusters  in redshift range $0.078 \leq z \leq 1.063$ {from the Ref. \cite{mantz} (see Fig. 1b)}. These authors measured the gas mass fraction in spherical shells at radii near $r_{2500}$\footnote{This radius is the one within which the mean cluster density is 2500 times the critical density of the Universe at the cluster's redshift.}, rather than integrated at all radii ($< r_{2500}$) as in previous works. The effect of this is to significantly reduce the corresponding theoretical uncertainty in the gas depletion from hydrodynamic simulations (see Fig. 6 in their paper and also \cite{pla,bat}).

 \item The Union2.1 compilation SNe Ia sample \cite{suzuki} formed by 580 SNe Ia data in the redshift range $0.015 \leq z \leq 1.4$ (see Fig. 1a), fitted using SALT2 \cite{guy2007}. All analysis and cuts were developed in a blind manner, i.e., with the cosmology hidden. In this point it is important to detail our methodology: we need SNe Ia and galaxy clusters at identical redshifts. Thus, for each galaxy cluster, we select SNe Ia with redshifts obeying the criterion $|z_{GC} - z_{SNe}| \leq 0.005$ and calculate the following weighted average for the SNe Ia data:
\begin{equation}
\begin{array}{l}
\bar{\mu}=\frac{\sum\left(\mu_{i}/\sigma^2_{\mu_{i}}\right)}{\sum1/\sigma^2_{\mu_{i}}} ,\hspace{0.5cm}
\sigma^2_{\bar{\mu}}=\frac{1}{\sum1/\sigma^2_{\mu_{i}}}.
\end{array}\label{eq:dlsigdl}
\end{equation}
Then, we end up with 40, 29 and 25 $\bar{\mu}_i$ and $\sigma^2_{\bar{\mu}_i}$ measurements when the sample from \cite{mantz}, \cite{bonamente} and \cite{fil} is considered, respectively. Following  \cite{suzuki} we added quadratically a 0.15 systematic error to each SNe Ia distance modulus error.
 
\item The $T_{CMB}(z)$ sample is composed by 36 points (see Fig. 1c). The data at low redshifts are from SZE observations \cite{luzzi} and at high redshifts from observations of spectral lines \cite{hurier}. In total, this represents 36 observations of the CMB temperature at redshifts between 0 and 2.5. We also use the estimation of the current CMB temperature $T_0 = 2.725 \pm 0.002$ K \cite{mather} from the CMB spectrum as estimated from the COBE satellite.
\end{itemize}

\begin{figure*}[htb]
\centering
\includegraphics[width=0.47\textwidth]{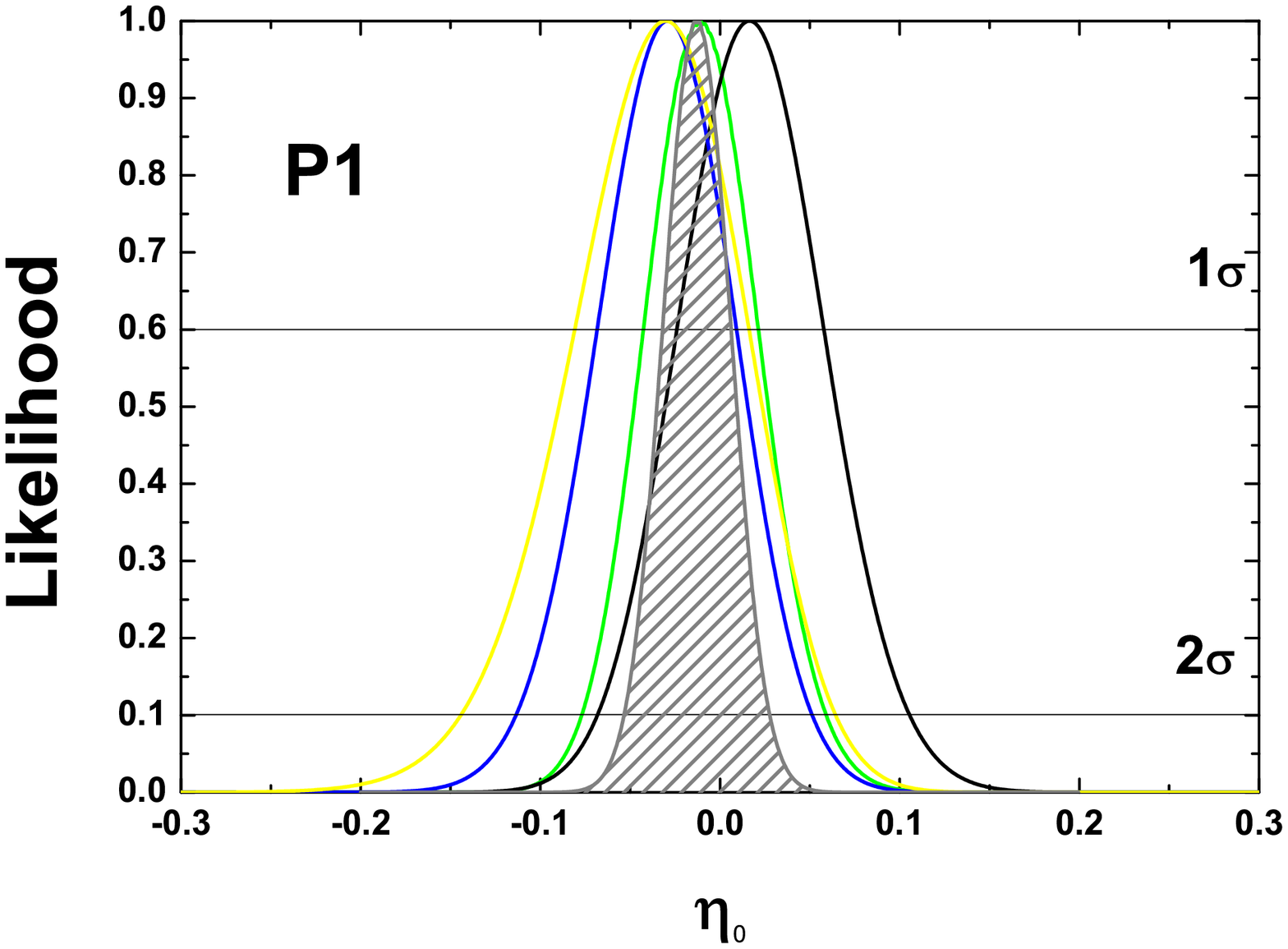}
\hspace{0.3cm}
\includegraphics[width=0.47\textwidth]{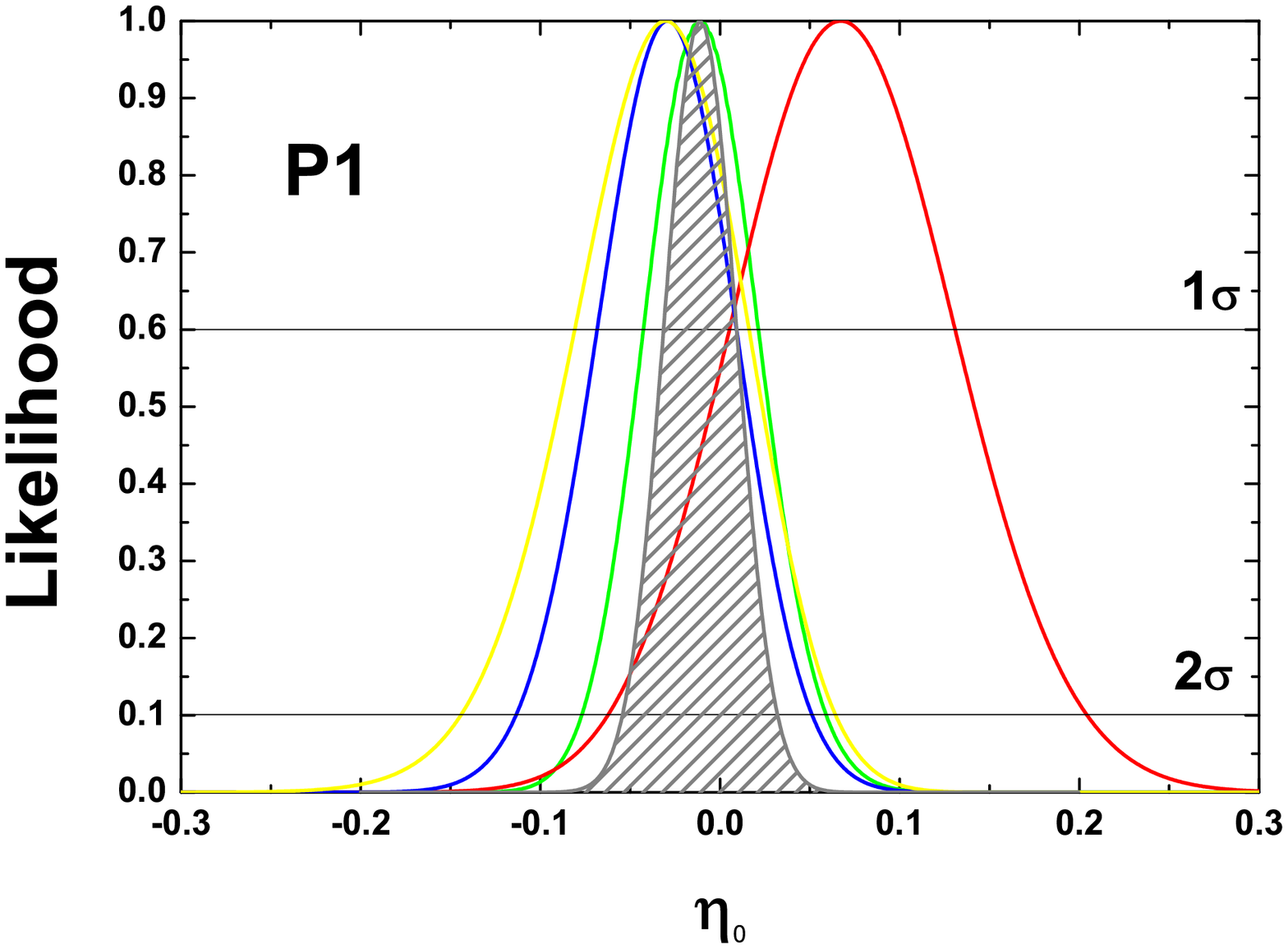} 
\\
\includegraphics[width=0.47\textwidth]{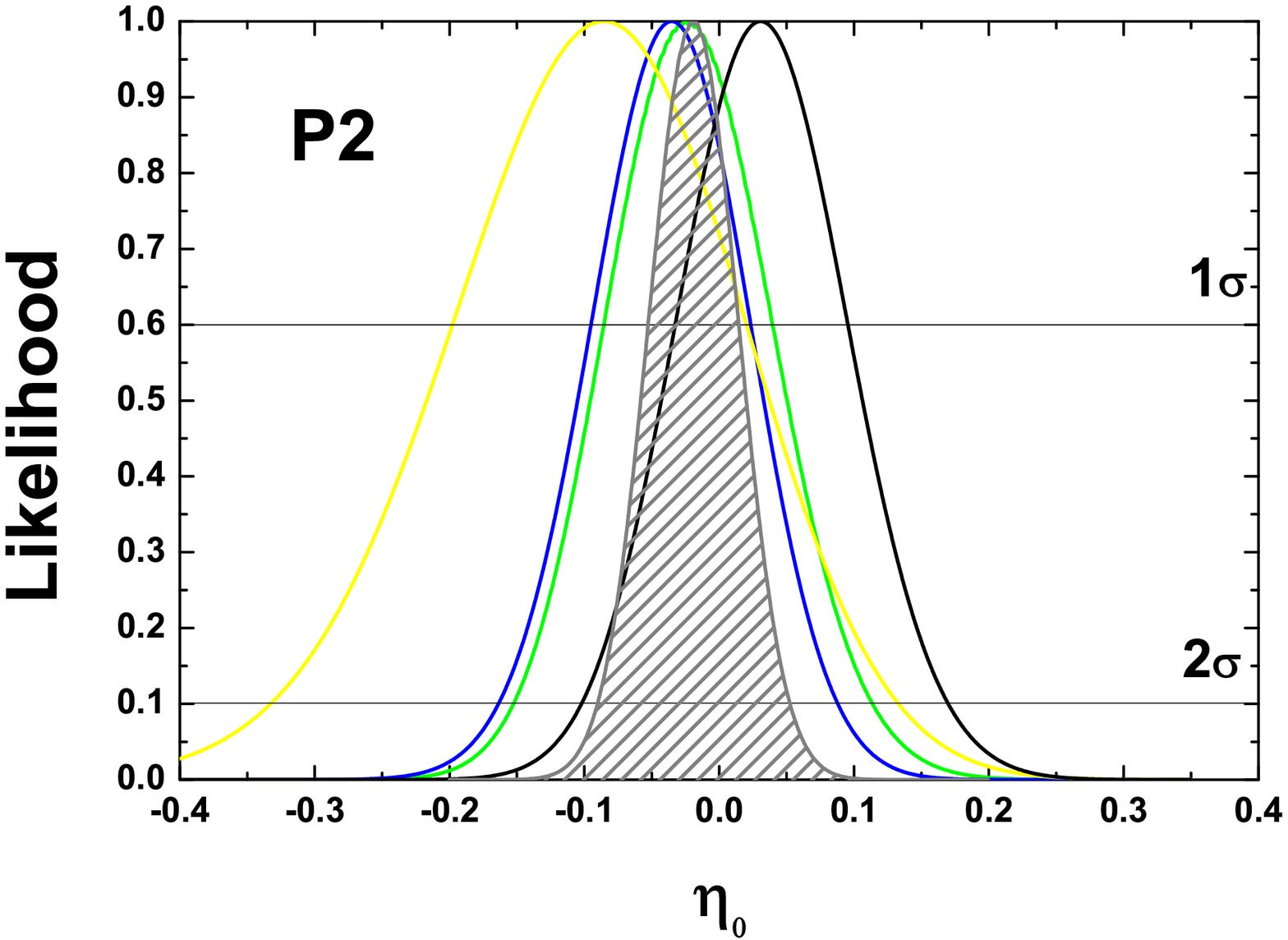}
\hspace{0.3cm}
\includegraphics[width=0.47\textwidth]{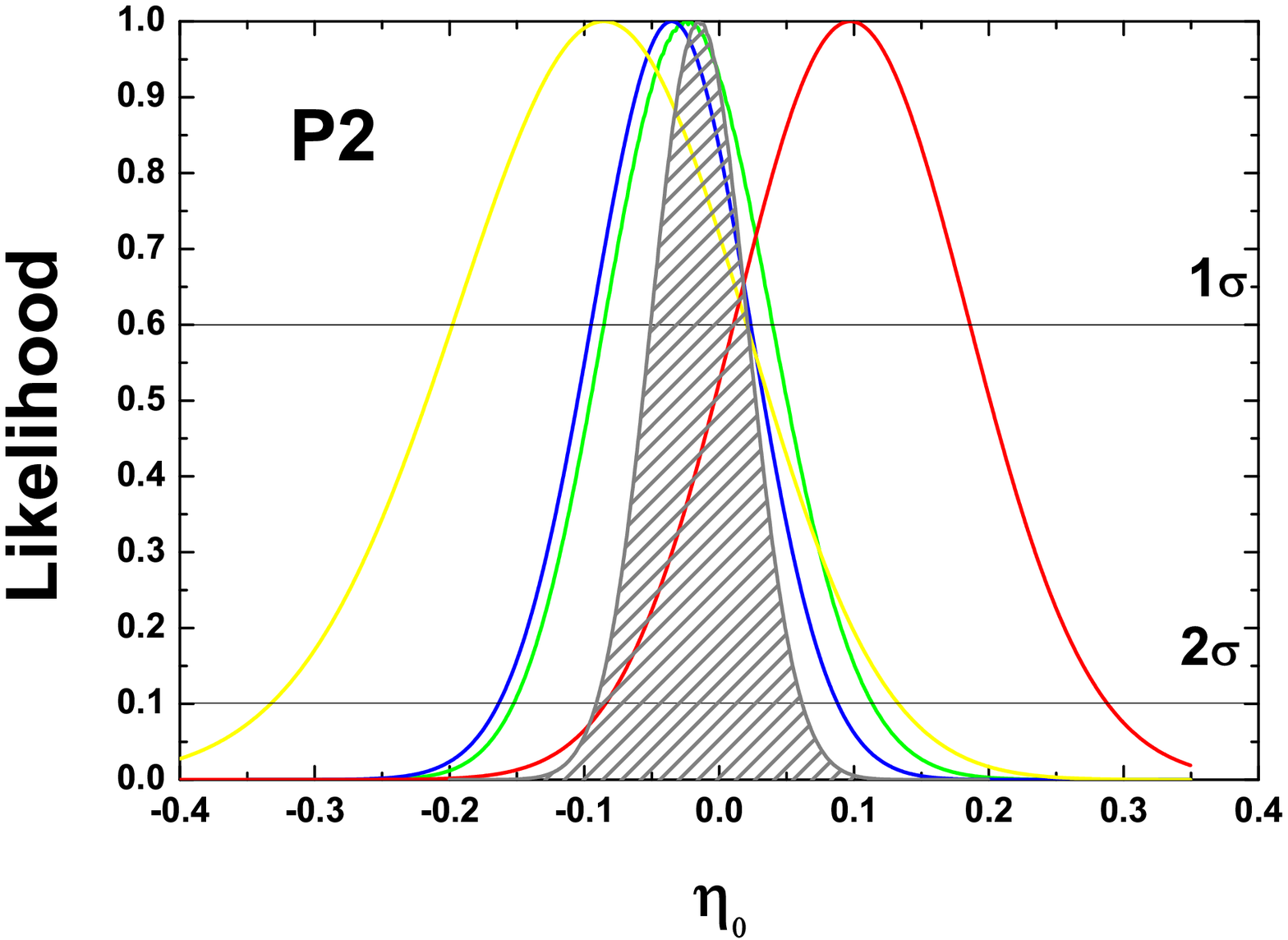}
\\
\includegraphics[width=0.47\textwidth]{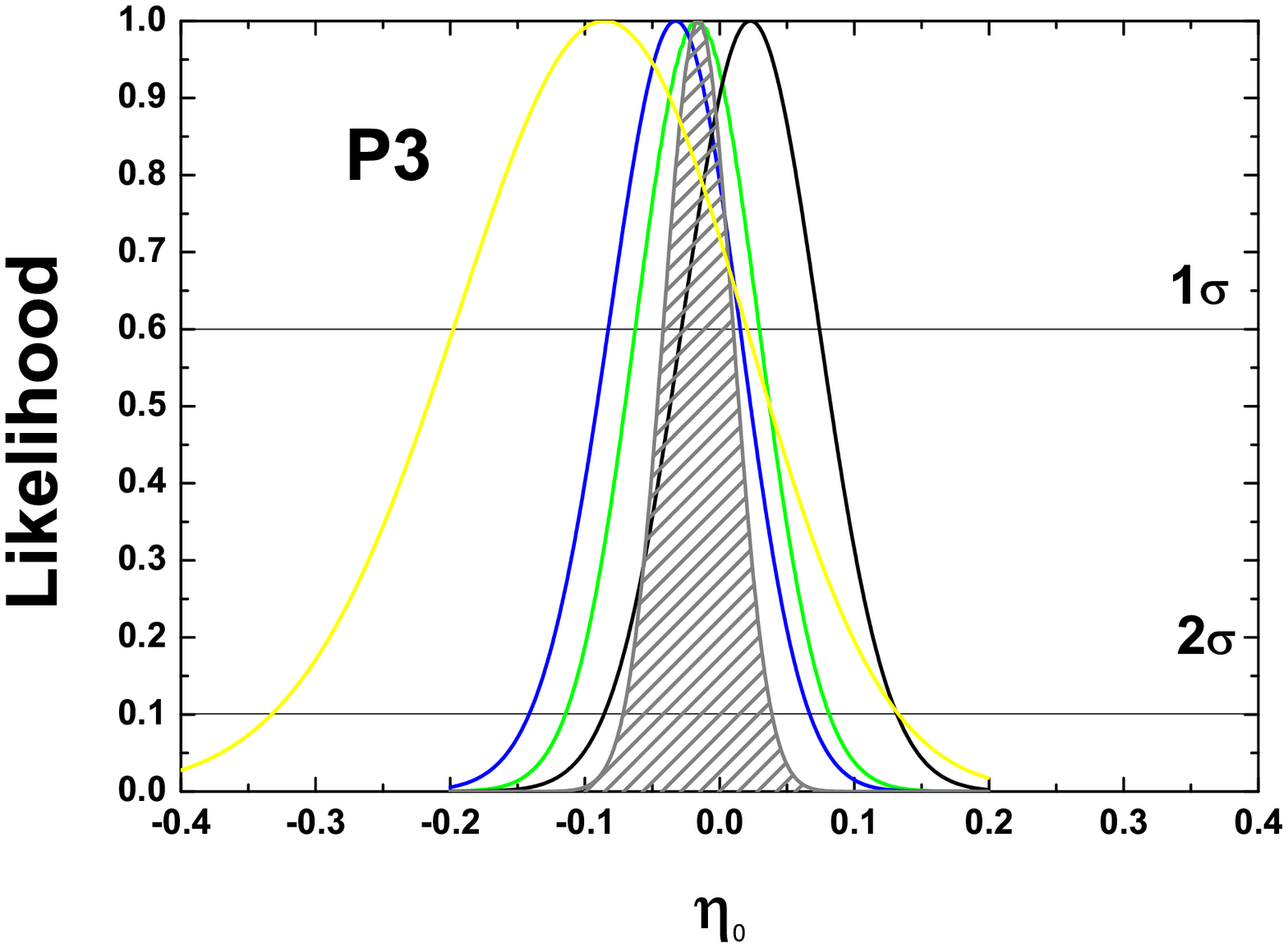}
\hspace{0.3cm}
\includegraphics[width=0.47\textwidth]{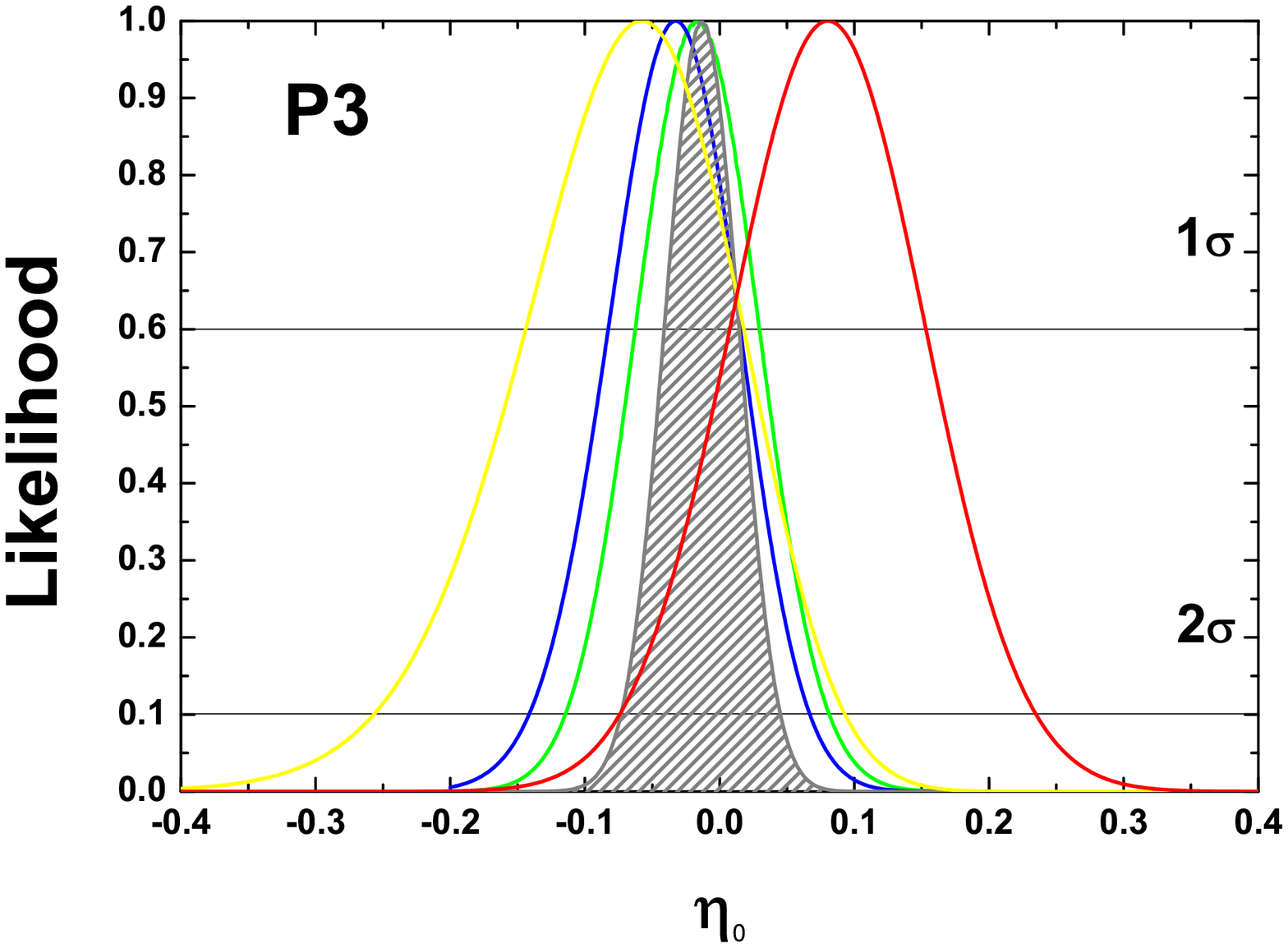}
\\
\includegraphics[width=0.47\textwidth]{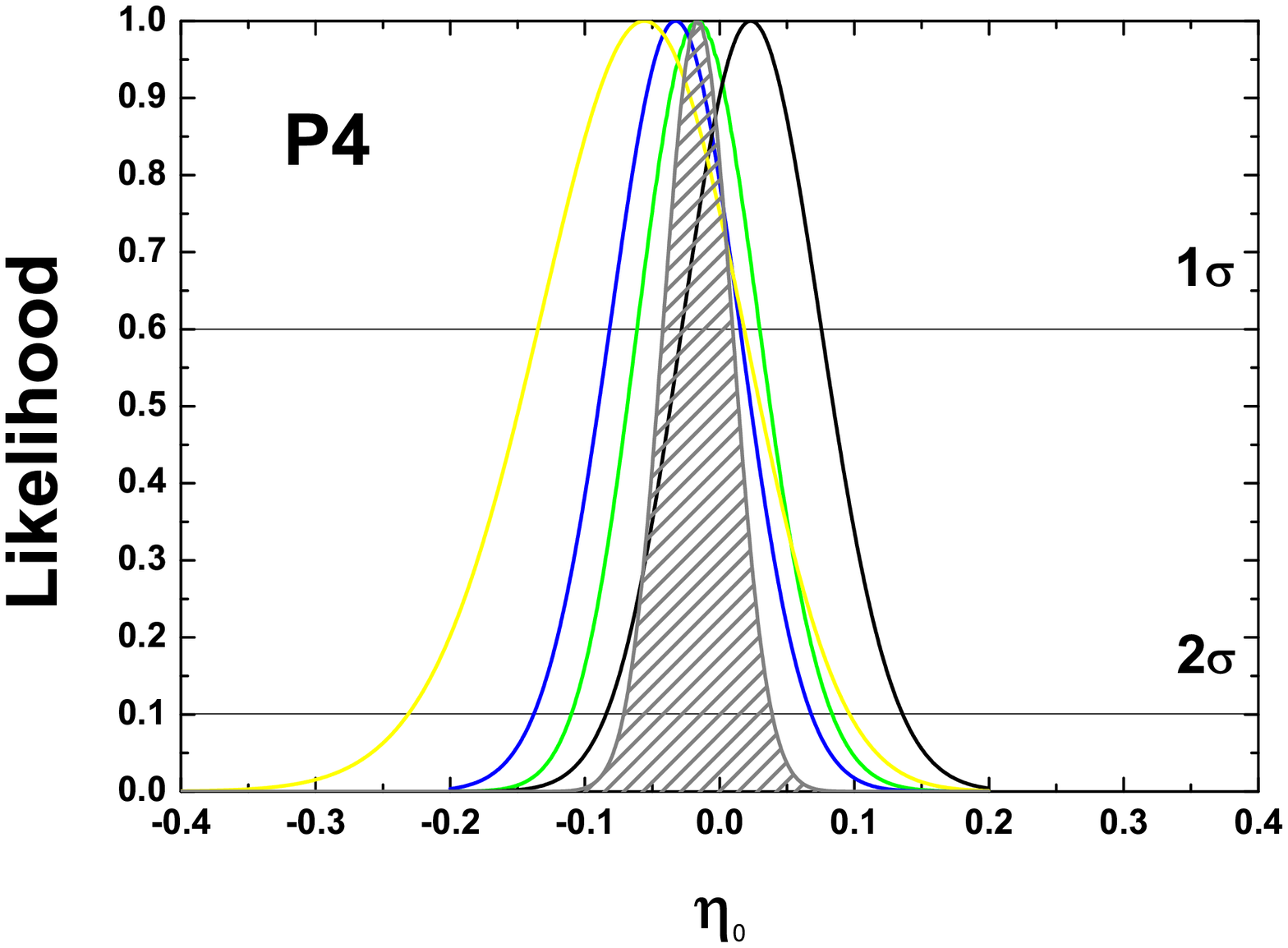}
\hspace{0.3cm}
\includegraphics[width=0.47\textwidth]{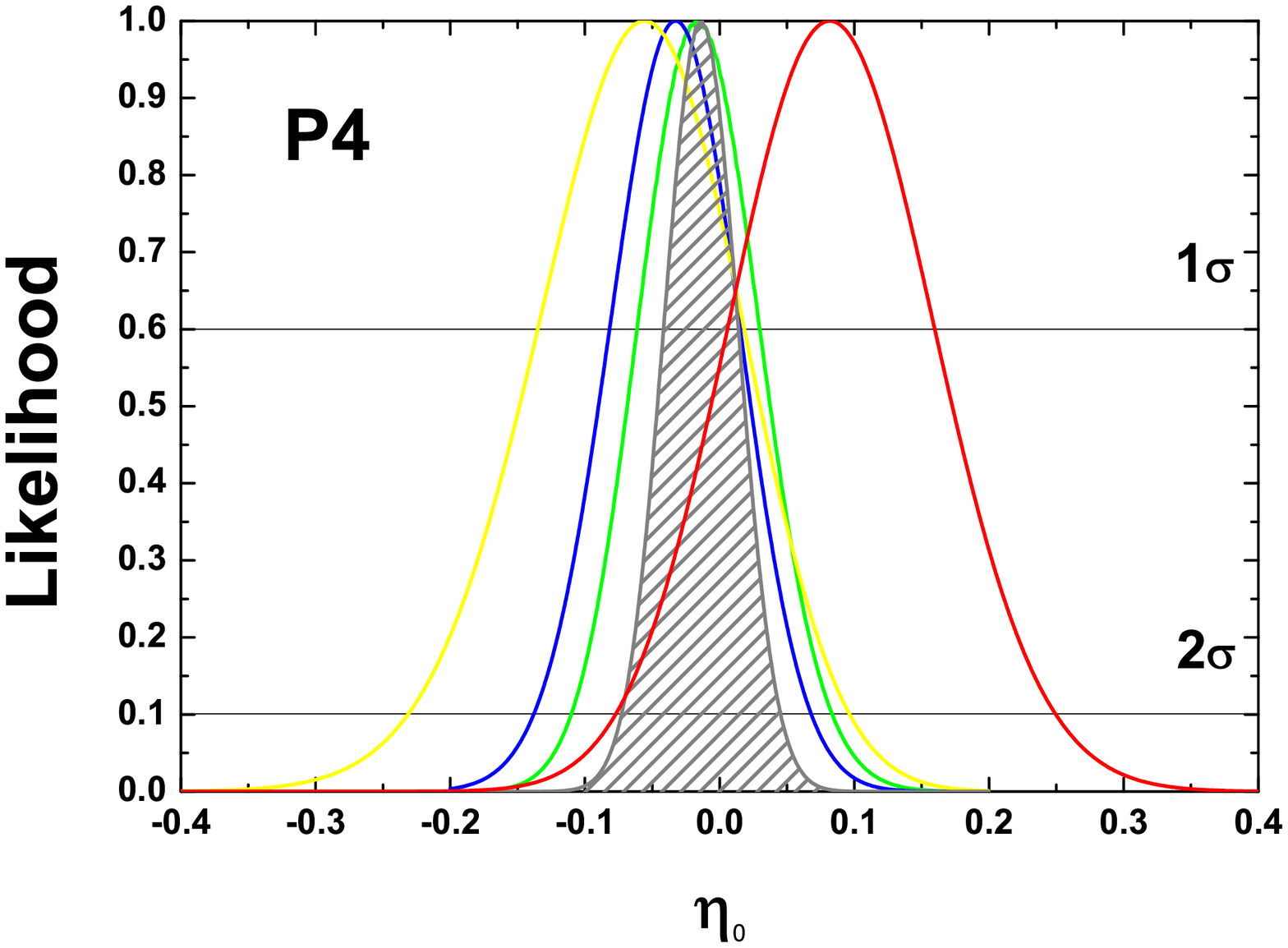}
\caption{Likelihood curves for the parametrizations P1, P2, P3 and P4, each row corresponds to a different parametrization. By considering the left panels, the solid green, black, blue and yellow lines correspond to {analyses} by using separately  GC1 + SNe Ia data,  GC2 + SNe Ia data,  $f^{obs}_{SZE}/f^{obs}_{X-ray}$ data and $T_{CMB}(z)$   in Eq. (\ref{chi}), respectively. On the other hand, by considering the right panels, the solid green, red, blue and yellow lines correspond to {analyses} by using GC1 + SNe Ia data,  GC3 + SNe Ia data,  $f_{SZE}/f_{X-ray}$ data, and $T_{CMB}(z)$, respectively. In all panels, the dashed area are the results of the joint analysis by using these data set.}
\end{figure*}

\section{Analises and Results}
 
We evaluate our statistical analysis by defining the likelihood distribution function, ${\cal{L}} \propto e^{-\chi^{2}/2}$, where 
\begin{eqnarray}
\chi^2 =\sum_{i=1}^{40}\frac{(\bar{\mu_1}(z_i)-\mu_{GC1}(\eta,z_i))^2}{\sigma_{obs}^2}
+\sum_{i=1}^{29}\frac{(\bar{\mu_2}(z_i)-\mu_{GC2}(\eta,z_i))^2}{\sigma_{obs}^2} 
+\sum_{i=1}^{25}\frac{(\bar{\mu_3}(z_i)-\mu_{GC3}(\eta,z_i))^2}{\sigma_{obs}^2} \nonumber \\
+\sum_{i=1}^{29}\frac{(\eta(z)^2 - f^{obs}_{SZE}/f^{obs}_{X-ray})^2}{\sigma_{obs}^2} %\nonumber  
 + \sum_{i = 1}^{38}\frac{{\left[ T(z_i) - T_{i,obs} \right] }^{2}}{\sigma^{2}_{T_i, obs}} ,
\label{chi}
\end{eqnarray} 
with GC1, GC2 and GC3 corresponding to samples from gas mass fraction \cite{mantz}, ADD \cite{bonamente} and ADD \cite{fil}, respectively, and $\sigma_{obs}^2= \sigma^2_{\bar{\mu}} + \sigma^2_{\mu GC}$ for each sample and  $T(z)$  given by Eq. (\ref{var}). The distance modulus $\bar{\mu_1}(i)$, $\bar{\mu_2}(i)$ and $\bar{\mu_3}(i)$ correspond to weighted averages from the SNe Ia data for each  i-galaxy cluster in samples present in Refs. \cite{mantz,bonamente,fil} (see Eq. 50). In our analysis, the normalization factor $N$ in Eq.(48) is taken as a nuisance parameter so that we marginalize over it. The EEP breaking is sought for allowing deviations from $\eta=1$ for parametrizations as (P1)-(P4), if  $\eta_0=0$  the {standard limit (with no interaction) for the electromagnetic sector is recovered}.  
 
The results for the parametrizations (P1), (P2), (P3) and (P4) are plotted in Figures 3, each row depicting a different parametrization. The left panels show the results with the $ADD$ data of Bonamente et al. \cite{bonamente}, while the right panels the $ADD$ data of de Filippis et al. \cite{fil}. By considering the left panels, the solid green, black, blue and yellow lines correspond to {analyses} by using separately  GC1 + SNe Ia data,  GC2 + SNe Ia data,  $f^{obs}_{SZE}/f^{obs}_{X-ray}$ data and $T_{CMB}(z)$   in Eq. (\ref{chi}), respectively. The dashed area are the results of the joint analysis by using these data set. On the other hand, by considering the right panels, the solid green, red, blue and yellow lines correspond to {analyses} by using GC1 + SNe Ia data,  GC3 + SNe Ia data,  $f_{SZE}/f_{X-ray}$ data, and $T_{CMB}(z)$, respectively. Again, the dashed area are the results of the joint analysis. As one may see from black and red lines, the results of the analyses by using ADD + SNe Ia data do not depend strongly on the galaxy cluster sample used. 

In Table I, we put our 1$\sigma$ results from the joint {analyses} for the four parametrizations and several $\eta_0$ values already present in literature which consider correctly possible variations of $\alpha$ and $\eta$ in their analyses. For completeness we also added the case $\eta(z)=\eta_0$ ($P_0$), in this case the standard limit (with no interaction) in the electromagnetic sector is recovered for $\eta_0=1$. As one may see, our results are in full agreement with each other and with the previous ones regardless the galaxy cluster observations and  $\eta(z)$ functions used. Moreover, our analysis presents most restrictive results and no significant  break of {EEP by means of the electromagnetic sector was verified}.

\begin{table*}[ht]
\caption{A summary of the current constraints on the parameters $c_0$ and $\eta_0$ for P0, P1, P2, P3 and P4,  from several cosmological data. The symbol * corresponds to angular diameter distance (ADD) from Ref. \cite{fil} and ** angular diameter distance from Ref. \cite{bonamente}.}
\label{tables1}%tab2
\par
\begin{center}
\begin{tabular}{|c||c|c|c|c|c|c|}
\hline\hline Reference & Data Sample & $\eta_0$ (P0) &$\eta_0$ (P1)& $\eta_0$ (P2)& $\eta_0$ (P3)& $\eta_0$ (P4)
\\ \hline\hline
\cite{holandaprd} & $ADD^{*}$+SNeIa      &- &$0.069 \pm 0.106$    & $0.000 \pm 0.135$ & - &- \\
\cite{holandasaulo} & $ADD^{**}$+SNeIa+$T_{CMB}$ &- & $-0.005 \pm 0.025$& $-0.048 \pm 0.053$ & $-0.005\pm 0.04 $&$-0.005 \pm 0.045$ \\
\cite{holandasaulo} & $ADD^{*}$+SNeIa+$T_{CMB}$ & -& $-0.005 \pm 0.032$& $-0.007 \pm 0.036$ &$ 0.015 \pm 0.045$ & $0.015 \pm 0.047$ \\
\cite{holandasaulo2} & GMF+SNeIa+$T_{CMB}$ & -& $-0.020 \pm 0.027$ & $-0.041 \pm 0.042$  & -  & - \\
\textbf{This paper} & $ADD^{**}$+GMF+SNeIa+$T_{CMB}$ & $1.006 \pm 0.010$ & $-0.012 \pm 0.022$ & $-0.02 \pm 0.034$  & $-0.017 \pm 0.026$  & $-0.017 \pm 0.027$ \\
\textbf{This paper} & $ADD^{*}$+GMF+SNeIa+$T_{CMB}$ & $1.005 \pm 0.010$ & $-0.011 \pm 0.021$ & $-0.015 \pm 0.033$  & $-0.013 \pm 0.028$  & $-0.013 \pm 0.027$ \\
\hline\hline
\end{tabular}
\end{center}
\end{table*}

\section{Conclusions}

The amount and quality of data gathered by cosmologists in the last decades allowed the establishment of a standard cosmological model, dubbed the flat $\Lambda$CDM model.
Along with it, these data also provide a myriad of opportunities to check the consistency of the cosmological framework and test for ideas beyond the standard model.
One of the fundamental hypotheses of the cosmological framework is the validity of the Einstein equivalence principle (EEP). As it was recently shown in Ref. \cite{hees}, a possible breakdown of the equivalence principle
in the electromagnetic sector can demonstrate distinct signatures, for instance, deformations of the cosmic distance duality relation (CDDR) and a time variation of the fine structure constant. 

In this paper, we have looked for possible deviations of the CDDR and a time-dependency of the fine structure constant as a test of the equivalence principle using four different observables at low and intermediate redshifts. The high complementarity of the samples due to their different degeneracies allowed us to improve constraints on the deviations of the CDDR between 20 and 40\%, depending on the parametrization adopted. The results point to a complete agreement with the validity of the EEP, which should be obeyed within a few percent regardless the considered parametrization. Future and ongoing surveys in different wavelengths will provide even more stringent tests to the EEP soon.

%%%%%%%%%%%%%%%%%%%%%%%%%%%%%%%%%%%%%%%%%%%%%%%%%%%%%%%%%%%%%%%%%%%%%%%
\begin{acknowledgements}
RFLH acknowledges financial support from  CNPq  (No. 303734/2014-0). SHP  acknowledges financial support from CNPq - Conselho Nacional de Desenvolvimento Cient\'ifico e Tecnol\'ogico, Brazilian research agency, for financial support, grants number 304297/2015-1 and 400924/2016-1. VCB is supported by S\~ao Paulo Research Foundation (FAPESP)/CAPES agreement under grant 2014/21098-1 and S\~ao Paulo Research Foundation under grant 2016/17271-5. CHGB is supported by CNPq under research project no 502029/2014-5.
\end{acknowledgements}
%%%%%%%%%%%%%%%%%%%%%%%%%%%%%%%%%%%%%%%%%%%%%%%%%%%%%
%%%%%%%%%%%%%%%%%%%%%%%%%%%%%%%%%%%%%%%%%%%%%%%%%%%%%%%%%%%%%%%%%%%%%%%%%%

\end{document}